# Control of electron-electron interaction in graphene by proximity screening

M. Kim[1], S. G. Xu[1,2], A. I. Berdyugin[1], A. Principi[1], S. Slizovskiy[1,2,3], N. Xin[1,2], P. Kumaravadivel[1,2], W. Kuang[1], M. Hamer[2], R. Krishna Kumar[1], R. V. Gorbachev[2], K. Watanabe[4], T. Taniguchi[4], I. V. Grigorieva[1], V. I. Fal'ko[1,2], M. Polini[5,1,6], A. K. Geim[1,2]

[1]School of Physics and Astronomy, University of Manchester, Manchester, M13 9PL, United Kingdom
[2]National Graphene Institute, University of Manchester, Manchester, M13 9PL, United Kingdom
[3]Saint-Petersburg INP, Gatchina 188300, Russia
[4]National Institute for Materials Science, Tsukuba, 305-0044, Japan
[5]Dipartimento di Fisica dell'Università di Pisa, Largo Bruno Pontecorvo 3, I-56127 Pisa, Italy
[6]Istituto Italiano di Tecnologia, Graphene Labs, Via Morego 30, 16163 Genova, Italy

**Electron-electron interactions play a critical role in many condensed matter phenomena[1-4], and it is tempting to find a way to control them by changing the interactions' strength. One possible approach is to place a studied electronic system in proximity of a metal, which induces additional screening and hence suppresses electron interactions. Here, using devices with atomically-thin gate dielectrics and atomically-flat metallic gates, we measure the electron-electron scattering length $\ell_{ee}$ in graphene at different concentrations $n$ and temperatures. The proximity screening is found to enhance $\ell_{ee}$ and change qualitatively its $n$ dependence. Counterintuitively, the screening becomes important only at gate dielectric thicknesses of a few nm, much smaller than the average separation $D \approx 1/\sqrt{n}$ between electrons. The critical thickness is given by ~$0.03\ \varepsilon D$, where $\varepsilon$ is the gate dielectric's permittivity, and the theoretical expression agrees well with our experiment. The work shows that, using van der Waals heterostructures with ultra-thin dielectrics, it is possible to modify many-body phenomena in adjacent electronic systems.**

Elementary electrostatics tells us that the electron charge $e$ placed at the distance $d$ from a bulk metal leads to a dipole potential evolving as $2ed^2/r^3$ at large in-plane distances $r \gg d$, which is much weaker than the original, unscreened Coulomb potential, $e/r$. Accordingly, a metallic gate placed sufficiently close to another electronic system can alter its electron-electron (e-e) interactions. Electrostatic screening by metallic gates has previously been employed to suppress charge inhomogeneity in graphene[5,6], alter its plasmon spectra[7,8] and renormalize an electronic structure of monolayer semiconductors[9]. In principle, proximity-gate screening may also affect e-e interactions. They can be parametrized by $\ell_{ee}$ and, a priori, it is unclear how close a metallic gate should be to change this parameter appreciably. From the above electrostatic considerations, one can infer that what matters most is the ratio $d/D$. For a two-dimensional (2D) electron system with typical $n = 10^{12}$ cm$^{-2}$, $D \approx 10$ nm and, therefore, the inferred gate separation $d \approx D$ is relatively easy to achieve experimentally. However, as shown below, the naïve expectations fail because of a small numerical factor $\delta$ such that e-e interactions for massless Dirac fermions are altered only if $d \leq \delta D \approx 0.03\ \varepsilon D$. For typical gate dielectrics with $\varepsilon < 5$, the required separation falls into a 1 nm range. For massive charge carriers such as those in bilayer graphene and 2D semiconductors, even smaller (atomic-scale) $d$ are necessary for efficient screening (Methods). It seems impossible to realize such small $d$ because of inevitable surface roughness of the metal and insulating films used for gating and electrical leakage through dielectrics of nanometer thickness. In this report, we achieve the extremely challenging conditions for proximity-gate screening by using van der Waals heterostructures with atomically-thin dielectric layers and atomically-flat gates.

Our devices were graphene monolayers encapsulated between hexagonal boron nitride (hBN) crystals whereas graphite monocrystals served as a bottom gate (Fig. 1). These heterostructures were fabricated using the standard dry-transfer procedures[5] described in Methods. Multiterminal Hall bar devices with several point contacts and closely placed voltage probes (Fig. 1a) were then defined by electron-beam lithography and plasma etching. An extra metal gate was deposited on top of the heterostructures, which allowed us to vary $n$ without applying voltages to the bottom screening gate. This was particularly important for our case of ultra-thin dielectrics to avoid their accidental breakdown and electrical leakage. The minimum thickness $d$ for the gate dielectric (Fig. 1b) was limited to 4 hBN layers (i.e. ~ 1.3 nm) because thinner crystals exhibited notable electron tunneling[10]. The devices typically had low-temperature ($T$) mobility $\mu$ of about $10^6$ cm$^2$ V$^{-1}$s$^{-1}$ and highly reproducible characteristics such that, at finite $T$, their longitudinal resistivity $\rho$ was practically independent of $d$ (Supplementary Fig. 1). This ensured that the reported behavior of $\ell_{ee}$ was due to changes in $d$ rather than transport characteristics. Because graphite is a semimetal with a relatively low carrier density of ~$10^{19}$ cm$^{-3}$, we also crosschecked that our conclusions were independent of the gate material using screening gates made from other layered metals such as Bi$_2$Sr$_2$CaCu$_2$O$_{8+x}$ and TaS$_2$ (Methods; Supplementary Fig. 2).

To demonstrate that e-e interactions can be tuned by proximity-gate screening, a reliable diagnostic tool is essential. Many quantum transport characteristics are known to be affected by the strength of e-e interactions. For example, the phase breaking length depends on it and can be measured in quantum interference experiments[11] (other possibilities are discussed in ref. 12). In principle, it should be possible to use such 'mesoscopic physics' tools to probe e-e interactions in graphene but, because of its ballistic transport at micrometer-scale distances, the approach is not easy to implement in practice and its results could be difficult to interpret. On the other hand, recent experiments have shown that graphene at finite $T$ and away from the charge neutrality point (NP) exhibits pronounced hydrodynamic effects[13-17], which allowed measurements of the kinematic electron viscosity $\nu_0$, and the extracted values of $\ell_{ee} = 4\nu_0/\nu_F$ were in quantitative agreement with theory ($\nu_F$ is the Fermi velocity). The viscosity measurements can be carried out using three complementary approaches: vicinity resistance[14,15], point contact geometry[16,18] and the viscous Hall effect[17]. Below we use all three to show that $\ell_{ee}$ changes with $d$. In another approach, we demonstrate that umklapp e-e scattering in graphene superlattices[19] is also affected by proximity-gate screening.

First, let us demonstrate the screening effect qualitatively. Figure 1c shows that the vicinity resistance $R_V$ is notably affected if a thin gate dielectric is employed. Vicinity measurements are discussed in detail in ref. 14 but, briefly, an electric current is injected through a narrow contact into a wide graphene channel. The negative voltage drop arising locally from a viscous electron flow is detected using a vicinity contact at a short distance $L$ from the current-injecting contact (Fig. 1a). One can see from Fig. 1c that, as $T$ increases, $R_V$ first decreases and then becomes negative. This indicates a transition from the ballistic transport regime (positive $R_V$) into a regime where ballistics is strongly affected by e-e scattering[15]. The minimum in $R_V(T)$ corresponds to the condition $\ell_{ee} \approx L$ and indicates an onset of hydrodynamic behavior[15]. As $\ell_{ee}$ decreases further with increasing $T$, $R_V$ becomes less negative and eventually positive, being dominated by currents caused by electron-phonon scattering[14,15]. The dependences $R_V(T)$ shown in Fig. 1c were measured for two similar devices at the same $L$. One had $d \approx 300$ nm (conventional Si back gate) whereas the other was made using 4-layer hBN as the gate dielectric. Despite the similar behavior of $R_V(T)$, the curve for $d \approx 1.3$ nm is clearly shifted to higher $T$. The shift direction indicates that the nearby gate caused an increase in $\ell_{ee}$, which is equivalent to a reduction in electron temperature by ~ 30 K. Note that, for $T$ above 100 K where the hydrodynamic regime develops, electron transport in high-quality graphene is universal and insensitive to experimental details.

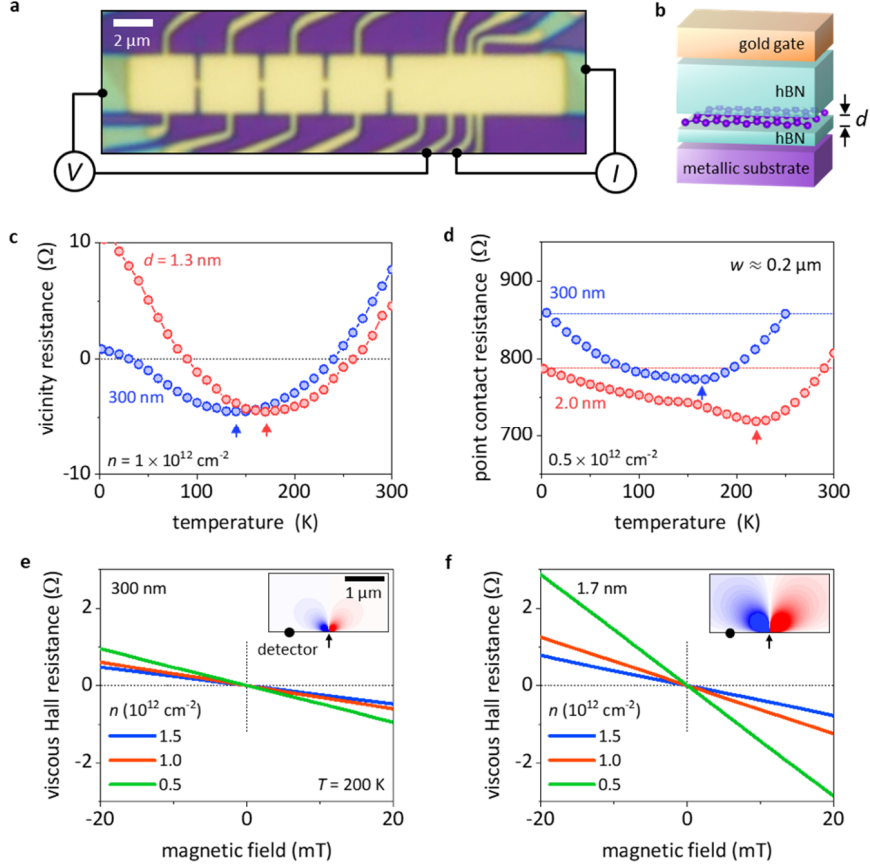

**Figure 1 | Graphene devices with proximity gating and its effect on electron hydrodynamics**. **a**, Optical micrograph of one of our devices with 4 sub-μm constrictions used for point-contact measurements and several closely spaced contacts for vicinity measurements. The wiring schematic illustrates current and voltage configurations for the latter measurements. **b**, Schematic side view of our heterostructures. **c**, $R_V$ as a function of $T$ for representative devices with a close graphite gate ($d \approx 1.3$ nm, red) and in the reference geometry ($d = 300$ nm, blue). The devices had similar geometry and $\mu$; same $L = 0.5$ μm. **d**, $R_{PC}(T)$ for screened and reference constrictions of the same width $w \approx 0.2$ μm (same color coding as in **c**). Dashed lines in **d** denote the resistance in the ballistic limit. Arrows in **c** and **d** indicate minima in $R_V$ and $R_{PC}$. **e** and **f**, Viscous Hall effect for reference and close-gate devices ($d = 300$ and 1.7 nm, respectively). The color-coded curves correspond to different $n$; all measurement conditions and geometries were same, including $L = 1$ μm and $T = 200$ K. The insets illustrate electric potentials that appear due to a viscous electron flow (the arrow and circle indicate positions of current and voltage contacts, respectively). The calculations[20] were carried out for the experimentally determined $\ell_{ee} \approx 0.3$ and $0.8$ μm for panels **e** and **f**, respectively; $B = 10$ mT. Blue-to-red color scale is arbitrary but same for both panels.

Similar phenomenology was observed in the point contact geometry (Fig. 1a). Again, the $T$-dependence of the point contact resistance $R_{\text{PC}}$ exhibits a clear minimum due to a viscous flow[16]. The shift to higher $T$ for the device with a proximity gate (Fig. 1d) indicates an increase in $\ell_{\text{ee}}$ for a given $T$. Such influence of the proximity gating was consistently observed in all our experiments. The $R_V$ and $R_{\text{PC}}$ dependences could also be used to extract $\ell_{\text{ee}}(T)$ following the recipe reported in refs. 14,16. Unfortunately, we found that, for atomically-thin gate dielectrics, detailed behavior of $R_V$ and, to some extent, $R_{\text{PC}}$ notably varied between different devices with nominally the same $d$. Those variations can be traced back to the fact that $R_V$ is sensitive to current injector's geometry[14] whereas a viscous contribution to $R_{\text{PC}}$ becomes smaller for close-gate devices as compared to those with thicker gate dielectrics.

In contrast to the vicinity and point-contact measurements, the viscous Hall effect[17] was found to be very robust, yielding quantitatively same results for different devices with same $d$. Accordingly, for quantitative analysis of how $\ell_{\text{ee}}$ depended on $d$, we focused on the latter measurements. The Hall viscosity experiments utilize the already discussed vicinity geometry (Fig. 1a) but a non-quantizing magnetic field $B$ is applied perpendicular to graphene[17]. The field leads to an asymmetry in the potential created by the viscous flow around the injection contact (insets of Figs. 1e,f). The viscous contribution asymmetric in $B$ is called the viscous Hall resistance $R_A$ and given by[17,20]

$$R_A = \rho \xi\left(\frac{L}{\sqrt{\nu_0 \tau}}\right) \frac{B}{B_0}, \qquad (1)$$

where $\xi(x)$ is a dimensionless function[20], $\tau$ is the transport scattering time, $B_0 = E_F/(8|e|\nu_0)$ is a characteristic magnetic field, and $E_F$ is the Fermi energy. Because $|\xi(x)|$ is a monotonically decreasing function of its argument for $x > 0$, $|R_A|$ increases with increasing $\ell_{\text{ee}}$ and, accordingly, devices with weaker e-e scattering should exhibit larger $|R_A|$.

To illustrate the effect of proximity-gate screening on Hall viscosity, Figs. 1e,f plot $R_A(B)$ for two representative devices with $d \approx 1.7$ and $300$ nm. The curves are taken under exactly the same conditions for several same $n$. As the two devices exhibited close $\rho$ and $\tau$ (Supplementary Fig. 1), the profound difference between Figs. 1e and 1f can only be attributed to different screening. The device with the thin dielectric exhibited much larger Hall viscosity than the reference device, and the effect was most pronounced at low $n$. This behavior proves again that the proximity screening suppresses e-e scattering, in agreement with the conclusions reached from the vicinity and point-contact measurements.

For the known transport characteristics ($\rho$ and $\tau$), Eq. 1 allows us to convert $R_A$ into $\ell_{\text{ee}}$, as described in detail in ref. 17. Figure 2a shows examples of $\ell_{\text{ee}}(T)$ found for close-gate and reference devices. At all $T$, the screened device displays $\ell_{\text{ee}}$ approximately twice longer than that in the standard device of the same electronic quality. This agrees well with many-body theory (solid curves in Fig. 2; Supplementary Fig. 3). Importantly, the proximity-gate screening qualitatively changes the dependence $\ell_{\text{ee}}(n)$ so that, away from the NP, $\ell_{\text{ee}}$ decreases with increasing $n$ (Fig. 2b). This contrasts with monotonically increasing $\ell_{\text{ee}}(n)$ for the reference devices, which was also reported previously[16,17]. Figure 2c summarizes our results by showing $\ell_{\text{ee}}$ measured for more than 10 different devices at characteristic $n$ and $T$ where viscous effects become most pronounced in graphene.

Despite the experimental scatter, Fig. 2c clearly shows that $\ell_{ee}$ can be altered appreciably by using thin gate dielectrics, if $d$ is smaller than a few nm.

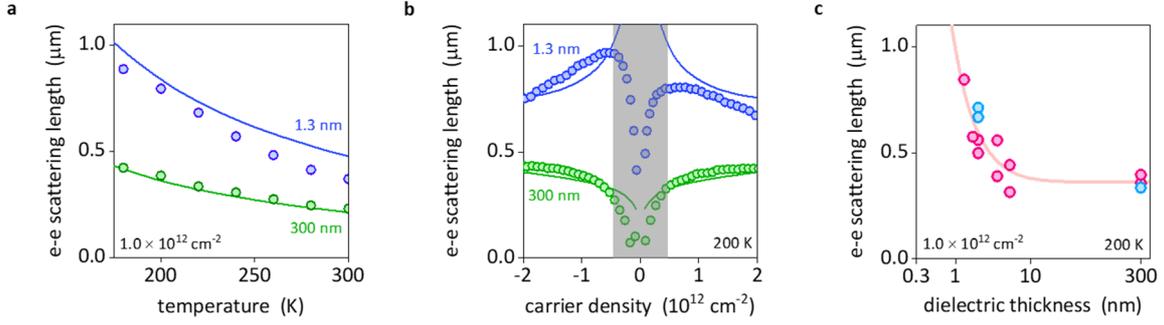

**Figure 2 | Dependence of the e-e scattering length on distance to the gate**. **a**, $\ell_{ee}(T)$ extracted from Hall viscosity measurements for the given $n$. Data for a close-gate device (blue symbols) are compared with a reference (green). **b**, Density dependence of $\ell_{ee}$ at 200 K (same color coding as in **a**). The grey-shaded region indicates the regime near the NP where the single-component hydrodynamic theory is not applicable[14,15,21] and, also, the cyclotron diameter became comparable with the width of our devices[17]. **c**, $\ell_{ee}$ as a function of $d$ for the given $n$ and $T$. Red and blue symbols: Results from Hall viscosity and point-contact measurements, respectively; shown are the average values for electron and hole doping (see panel **b** for an example of scatter due to electron-hole asymmetry). For all the panels, the solid curves are theoretical results (Supplementary Information).

To explain the observed dependences of $\ell_{ee}$ on $n$ and $d$, we carried out numerical calculations in the random phase approximation for the dynamically screened interactions[12,22,23]. The metallic gate was modelled as a perfect conductor, and small departures from this model caused by a finite carrier density were estimated in Supplementary Section 4. The results are shown by the solid curves in Fig. 2. No fitting parameters were used, except for multiplying all the theoretical curves by the same small factor of 1.3 (its non-Fermi-liquid origins are discussed in Supplementary Section 4). However, to gain better insight about the observed behavior, we also derived the following analytical expression

$$\ell_{ee} \approx \frac{4\hbar v_F E_F}{\pi} \frac{1}{(k_B T)^2 \ln\left(\frac{2E_F}{k_B T}\right)} \left(\frac{1+2dq_{TF}}{2dq_{TF}}\right)^2, \qquad (2)$$

where $k_F = \sqrt{\pi n}$ and $q_{TF} = 4\alpha_{ee} k_F$ are the Fermi and Thomas-Fermi wavenumbers, respectively. Here, $\alpha_{ee} \approx 2.2/\varepsilon$ is graphene's coupling constant and $k_B$ is the Boltzmann constant (Supplementary Section 4 discusses the case of generally anisotropic $\varepsilon$). The expression is accurate in the Fermi-liquid regime ($k_B T \ll E_F$), where it matches our numerical results (Supplementary Section 4). The last term in Eq. 2 appears due to the gate presence, and the key parameter describing its screening effect is $dq_{TF}$. In the far-gate regime, $d \gg 1/q_{TF}$, Eq. 2 reduces to the standard unscreened expression[23]. In the opposite limit, $d \ll 1/q_{TF}$, e-e scattering is strongly reduced due to screening, and $\ell_{ee}$ increases with decreasing both $d$ and $n$, as $1/d^2$ and approximately $1/\sqrt{n}$, respectively, in agreement with our experiment (Fig. 2). The latter dependence is opposite to the unscreened case, where $\ell_{ee}$ increases as $\sqrt{n}$, in agreement with the results of Fig. 2b. The crossover between the far- and close- gate regimes occurs at a critical distance $d_c$ such that $d_c \approx 1/2q_{TF} = 1/(8\alpha_{ee} k_F)$, which translates into the previously introduced parameter $\delta \approx 0.03\varepsilon$. For hBN with $\varepsilon \approx 3.5$ and at typical $n = 10^{12}$ cm$^{-2}$, we obtain $d_c \approx 1.1$ nm, which explains why the gate screening becomes noticeable only for our smallest $d$ (Fig. 2c). Further information about our theoretical analysis is provided in Supplementary Information.

To check how robust our conclusions are, we have also examined the effect of gate-induced screening on umklapp e-e scattering[19] that dominates resistivity $\rho$ of graphene-on-hBN superlattices at elevated $T$. We made several superlattice devices with the moiré periodicity $\lambda \approx 15$ nm, as confirmed by the periodicity of Brown-Zak oscillations[24] and the appearance of secondary NPs[25-28] at the expected $n$ (Fig. 3a). One of the devices was the standard Hall bar with $d = 300$ nm, like those reported previously[19]. The other two were same in design but had a bottom graphite gate placed at short $d$, as in the above viscosity experiments. Figure 3 shows typical $\rho(n,T)$ measured for these graphene superlattices. For $d = 300$ nm, the observed behavior was same as reported previously, and the $T$ dependent part ($\Delta\rho$) of graphene superlattice's resistivity could be described quantitatively by umklapp e-e scattering[19]. It is responsible for the rapid increase of $\Delta\rho \propto T^2$ (Fig. 3b). The proximity-gate screening notably suppressed $\Delta\rho(T)$, by a factor $> 2$ for $d \approx 1.3$ nm. Our theoretical analysis (Supplementary Section 5) shows that $\Delta\rho$ for the close-gate devices should exhibit the same $T$ dependence ($\propto T^2$) but with a reduced absolute value. The umklapp e-e scattering length, $\ell_{ee}^U$, is governed by distinctive processes with a momentum transfer of $\sim \hbar g$ where $g = \frac{4\pi}{\sqrt{3}}\lambda^{-1}$ is the superlattice reciprocal vector. As shown in Supplementary Section 5, proximity screening for $\ell_{ee}^U$ becomes important if $d < 0.1\lambda$, which again means that few-nm-thick gate dielectrics are essential to observe the screening effect. It is convenient to quantify this effect by the dimensionless ratio, $\Delta\rho(\infty)/\Delta\rho(d) \approx \ell_{ee}^U(d)/\ell_{ee}^U(\infty)$. The results are plotted in the inset of Fig. 3b and show good agreement with theory (for details, see Supplementary Section 5).

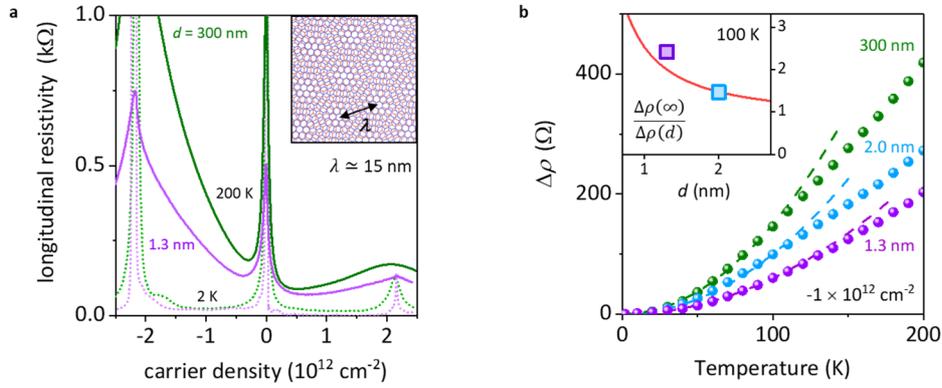

**Figure 3 | Suppression of umklapp e-e scattering in graphene superlattices by proximity-gate screening. a,** $\rho(n)$ of graphene-on-hBN superlattices for $d \approx 1.3$ and 300 nm (purple and green curves, respectively). Dotted and solid curves: $T = 2$ and 200 K, respectively. Inset: Illustration of the moiré pattern arising from crystallographic alignment between graphene and hBN lattices. **b,** $T$-dependent part of $\rho$ for superlattice devices with different $d$ (color-coded symbols); $n = -1 \times 10^{12}$ cm$^{-2}$ so that superlattices' first Brillouin zones are approximately half-filled with holes[26-28]. Dashed curves: Best fit to the predicted $T^2$ dependence[19]. All the devices had $\lambda \approx 15$ nm and close $\rho$ at 2 K. Inset: $\Delta\rho(d)$ for the two close-gate superlattices normalized by $\Delta\rho(\infty)$ measured for the reference (far-gate) superlattice. The color-coded symbols in the inset are taken from the main panel and valid for all $T \leq 120$ K because of the $T^2$ dependence. Solid curve: Theory.

To conclude, e-e scattering in monolayer graphene at finite $n$ can be strongly suppressed if a metallic gate is placed at $d$ of $\sim 1$ nm. This "close-gate" regime has become accessible due to the use of van der Waals assembly that allows atomically sharp interfaces and ultra-thin dielectrics. It is tempting to exploit the outlined strategy to assess interaction phenomena near the NP where low $n$ allow the condition $d \ll 1/\sqrt{n}$ to be satisfied easier but interpretation of some observations had proven difficult. Other interesting candidates are exotic phenomena driven by strong correlations (e.g., various many-body phases in twisted bilayer graphene[29,30]) and, especially, interaction effects governed by lengths longer than $\ell_{ee}$. The experimental challenge to reach the close-gate regime can partially be mitigated by using high-$\varepsilon$ dielectrics.

**Methods**

**Device fabrication.** Our heterostructures were assembled using 'stamps' made from polypropylene carbonate (PPC) as a sacrificial polymer placed on polydimethylsiloxane (PDMS). Such polymer stamps were used to pick up exfoliated thin crystals in the following sequence: top hBN (typically thicker than 30 nm), monolayer graphene and thin bottom hBN. The latter served as a gate dielectric in the final device configuration (Fig. 1b), and its thickness was determined by atomic force microscopy. The resulting hBN/graphene/hBN stack was then released onto relatively small graphite crystals with thickness of $3-10$ nm, which were prepared in advance on an oxidized Si wafer. The stack was large enough to extend outside the bottom graphite region, which allowed us to make quasi-one-dimensional contacts to graphene[31] without electrically contacting the graphite gate. The metallic contacts were defined by electron-beam lithography. We first used a mixture of $CHF_3$ and $O_2$ to plasma-etch hBN/graphene and expose the required contact regions. This was followed by deposition of 2 nm Cr/ 60 nm Au to make Ohmic contacts to graphene. A gold top gate was then fabricated using another round of electron-beam lithography and, also, served as an etching mask for the final etching step to define the Hall bar geometry.

The devices with other metallic gates ($Bi_2Sr_2CaCu_2O_{8+x}$ and $TaS_2$) required fabrication in an oxygen- and moisture- free atmosphere of a glovebox[32] to avoid deterioration of the metal surfaces. Even using glovebox encapsulation, we observed a notable reduction in graphene's quality for the above gate materials, presumably because of electrical charges at the exposed surfaces (for small $d$, typical $\mu$ became $< 10^5$ $cm^2$ $V^{-1}s^{-1}$ and charge inhomogeneity near the NP considerably increased). Accordingly, reliable measurements of $\ell_{ee}$ in this case were only possible at high $n \gtrsim 2.0 \times 10^{12}$ $cm^{-2}$ (Supplementary Section 2). We also note that encapsulated graphene devices with the conventional gates made by metal deposition on top of a thin gate dielectric ($d < 2$ nm) exhibited extremely low $\mu$ of only $\sim 10^4$ $cm^2$ $V^{-1}s^{-1}$. Such poor electronic quality made it impossible to carry out the $\ell_{ee}$ measurements described in the main text.

**Electrical measurements.** The devices were measured in a variable temperature insert that allowed stable $T$ between 2 and 300 K. The standard lock-in amplifier techniques were employed using excitation currents of typically $0.1-1$ μA at a frequency of 30.5 Hz. For measurements of Hall viscosity, we used the same vicinity geometry as shown in the schematic of Fig. 1a. The distance between injector and detector contacts was usually between 0.5 and 1.5 μm. The viscous Hall resistance was determined as an antisymmetric-in-$B$ component of the vicinity resistance in fields below ±30 mT. For the point-contact measurements, we employed the quasi-four-probe geometry by driving the current through the wide contacts (on the left and right in Fig. 1a) and using the leads next to the studied constrictions as voltage probes.

**Proximity screening for systems with the parabolic spectrum**. The close-gate condition depends on the density of states at the Fermi energy of the material one wants to control. We have studied graphene not only because of its electronic quality but also because of the low-density of states provided by its Dirac spectrum. For a 2D system with the conventional parabolic spectrum, the close-gate condition is much more difficult to achieve. In the latter case, a proximity metal gate can provide efficient screening of e-e interactions only for distances $d$ below $\sim \varepsilon\, m_e a_B\, /(2\, N_f\, m_{\text{eff}})$, where $a_B \approx 0.5$ Å is the Bohr radius, $m_e$ and $m_{\text{eff}}$ are the free-electron and effective masses, respectively, and $N_f$ is the number of spin/valley flavors. Here, $\varepsilon = \varepsilon_\perp$ is the perpendicular component of the dielectric permittivity of a gate dielectric. For bilayer graphene[33,34] with $N_f = 4$, $m_{\text{eff}} \geq 0.03\, m_e$ and using hBN as a dielectric ($\varepsilon_\perp \approx 3.5$), the close-gate condition requires $d < 7$ Å, which is essentially out of experimental reach.

**Acknowledgements**

This work was supported by the European Research Council, Graphene Flagship, EPSRC Grand Challenges (EP/N010345/1), the Royal Society and Lloyd's Register Foundation. A.I.B., W.K. were supported by Graphene NowNANO Doctoral Training Programme.

## #1. Mean free path and mobility

We carefully examined transport characteristics for several monolayer graphene devices with different dielectric thicknesses $d$. The mean free path $\ell$ with respect to momentum-non-conserving collisions was determined from the measured longitudinal resistivity $\rho$ by using the Drude formula. The carrier density $n$ was found from Hall measurements. Typical results for $\ell$ as a function of $n$ are shown in Supplementary Fig. 1a. The mean free path first increases with increasing $n$ and then saturates for $n \gtrsim 1.0 \times 10^{12}$ cm$^{-2}$. It monotonically decreases with temperature $T$ as expected. Such behavior was observed for all the measured devices independently of their $d$. This is elucidated by Supplementary Fig. 1b that shows $\ell$ for different $d$ at the given $n$ at room $T$. One can see that the measured $\ell$ varied only slightly, from ~ 0.7 to 1.1 μm, depending on graphene device's quality. Similarly, carrier mobilities $\mu(n)$ exhibited little dependence on $d$ (Supplementary Fig. 1c).

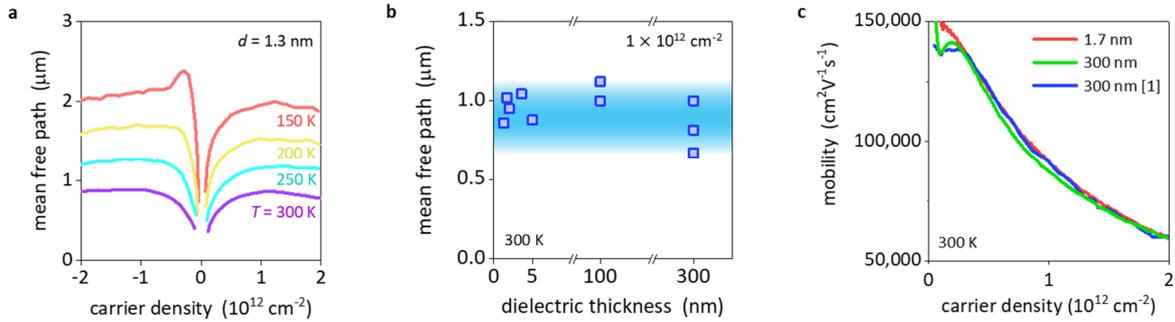

**Supplementary Figure 1 | Transport characteristics for different thicknesses of the gate dielectric. a,** $\ell(n)$ for a graphene device with $d \approx 1.3$ nm at a few representative $T$. **b,** $\ell$ for devices with different $d$ at 300 K; $n = 1 \times 10^{12}$ cm$^{-2}$. **c,** Density dependence $\mu(n)$ at room $T$. The mobilities measured for devices with different $d$ collapse on a single curve. The red and green curves are for gate dielectrics with ≈ 1.7 and 300 nm, respectively. The blue curve: Data from ref. 1 to indicate the generality of such behavior at elevated $T$.

## #2. Different screening materials

Because graphite is a semimetal[2,3] with a relatively low carrier concentration of the order of $10^{19}$ cm$^{-3}$, we have checked the generality of our conclusions using other metallic substrates, namely Bi$_2$Sr$_2$CaCu$_2$O$_{8+x}$ (BSCCO) and TaS$_2$ which have concentrations of ~ $10^{22}$ cm$^{-3}$ (ref. 4). To this end, devices similar to those shown in Fig. 1a of the main text were fabricated but, instead of graphite, cleaved BSCCO and TaS$_2$ crystals served as metallic substrates. To protect them from degradation, fabrication had to be carried out in an argon atmosphere of a glovebox as discussed in Methods. The carrier mobility $\mu$ for the latter devices was comparable to that of the devices made with graphite screening gates but only for high $n \gtrsim 2 \times 10^{12}$ cm$^{-2}$. At lower $n$, the electronic quality was insufficient to probe electron viscosity because of short $\ell$, presumably due to extra charges that appear on the metallic surfaces exposed to the ambient atmosphere. Accordingly, for the alternative screening substrates, we worked in the high $n$ regime to measure the viscous Hall resistance and then extract $\ell_{ee}$. Supplementary Fig. 2 shows the resulting $\ell_{ee}$ for graphene devices using various screening materials. Within our experimental accuracy, no difference in $\ell_{ee}$ could be noticed, and the experimental data closely followed the theoretical predictions.

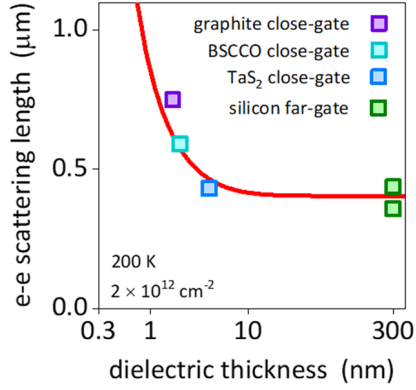

**Supplementary Figure 2 | Electron-electron scattering in devices with different materials used for proximity screening.** Symbols: Measured $\ell_{ee}$ at 200 K and $2 \times 10^{12}$ cm$^{-2}$ (color coded). Solid curve: Theory.

**#3. Point contact geometry**

For completeness, we also measured $\ell_{ee}$ using the point-contact geometry[5]. By applying an electric current through a graphene constriction and monitoring a voltage drop at nearby contacts (see Fig. 1a of the main text), the point contact resistance $R_{PC}$ was measured. Supplementary Fig. 3a shows $R_{PC}(T)$ for a graphene constriction with a geometrical width of $\sim 350$ nm as found by atomic force microscopy. The transport width $w$ of the constriction was somewhat smaller, $\sim 270$ nm, as found by fitting $R_{PC}(n)$ at liquid-helium $T$ by the standard Sharvin formula ($R_{Sh} = \frac{\pi h}{4e^2} \frac{1}{w\sqrt{n\pi}}$). The smaller width inferred from the fit is expected and presumably caused by edge roughness[5]. $R_{PC}$ exhibited a nonmonotonic $T$ dependence, becoming at intermediate $T$ notably smaller than the ideal value in the ballistic limit (Supplementary Fig. 3a). This "superballistic" behavior is due to e-e scattering as discussed elsewhere[5,6].

To extract $\ell_{ee}$ from the measurements such as those shown in Supplementary Fig. 3a, we used the expression[5,6]
$$R_{PC} = (1/R_{Sh} + G_v)^{-1} + R_C$$
where $R_C = b\rho$ is the contact resistance arising from the wide regions near the point contact. $R_C$ can be determined accurately for the known $\rho$ whereas the dimensionless coefficient $b$ is found from numerical simulations[5]. The viscous contribution $G_v$ to the point-contact conductivity is given by[6] $G_v = \frac{\sqrt{|n|\pi} e^2 w^2}{8\hbar \ell_{ee}}$. Supplementary Fig. 3b shows examples of $\ell_{ee}(T)$ found using the above analysis. The behavior of $\ell_{ee}$ agrees well with that found from the Hall viscosity measurements in the main text. For example, $\ell_{ee}$ is clearly enhanced for devices with close metallic gates. The experimental data also agree with theory whereas relatively small deviations from it at high $T$ are due to non-Fermi-liquid corrections as reported in ref. 5 and, also, explained below.

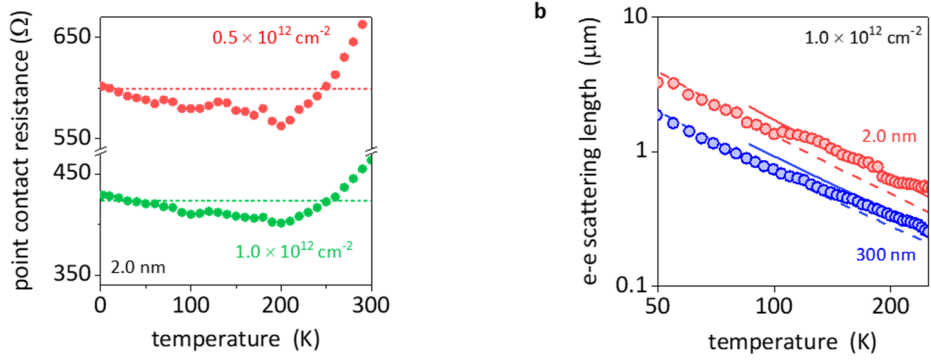

**Supplementary Figure 3 | Electron-electron scattering length found from point-contact measurements. a,** Point-contact resistance for a device with a close gate ($d \approx 2.0$ nm) at different $n$ (color coded). Dots: Experimental data. The dashed lines indicate the ideal value expected in the ballistic limit at low $T$. **b,** $\ell_{\rm ee}(T)$ for $d = 2.0$ (red) and 300 nm (blue) for the given $n$. Symbols: Experiment. Dashed curves: Theoretical predictions with no fitting parameters. Solid curves: Same theory data but multiplied by a numerical coefficient of 1.3.

#### #4. Microscopic theory of screened electron-electron scattering

In this Section we briefly described our approach to calculate $\ell_{\rm ee} = v_{\rm F}\tau_{\rm ee}$. The mean free time $\tau_{\rm ee}$ for e-e scattering is controlled by the one-body Green's function $G_\lambda(\boldsymbol{k}, \omega)$, where $\lambda = \pm 1$ is a band index ($\lambda = +1$ for conduction-band states and $\lambda = -1$ for valence-band states). This quantity satisfies the Dyson equation (setting $\hbar = 1$), $G_\lambda(\boldsymbol{k}, \omega) = [\omega - \xi_{\boldsymbol{k},\lambda} - \Sigma_\lambda(\boldsymbol{k}, \omega)]^{-1}$, where $\xi_{\boldsymbol{k},\lambda}$ are single-particle band energies measured from the chemical potential $\mu$ and $\Sigma_\lambda(\boldsymbol{k}, \omega)$ is the retarded self-energy. The latter quantity needs to be approximated. In weakly-correlated materials, a good approximation is the so-called $GW$ approximation[7,8] in which the electron self-energy is expanded to first order in the dynamically screened Coulomb interaction $W(\boldsymbol{q}, i\Omega)$

$$\Sigma_\lambda(\boldsymbol{k}, i\omega_n) = -k_{\rm B}T \sum_{\lambda'=\pm 1} \int \frac{d^2q}{(2\pi)^2} \sum_{m=-\infty}^{+\infty} W(\boldsymbol{q}, i\Omega_m) F_{\lambda\lambda'}(\theta_{\boldsymbol{k},\boldsymbol{k}-\boldsymbol{q}}) G_{\lambda'}(\boldsymbol{k} - \boldsymbol{q}, i\omega_n + i\Omega_m) \quad \text{(S1)}$$

where $\omega_n = (2n+1)\pi k_B T$ is a fermionic Matsubara frequency, the sum runs over all the bosonic Matsubara frequencies $\Omega_m = 2m\pi k_B T$, $\theta_{\boldsymbol{k},\boldsymbol{k}-\boldsymbol{q}}$ is the angle between $\boldsymbol{k}$ and $\boldsymbol{k} - \boldsymbol{q}$, and $F_{\lambda\lambda'}(\varphi) = [1 + \lambda\lambda' \cos(\varphi)]/2$ is the so-called chirality factor[9]. The retarded self-energy can be obtained after analytical continuation $i\omega_n \to \omega + i0^+$. For the sake of concreteness and without loss of generality due to particle-hole symmetry, we focus on electron-doped graphene, i.e. on the case $E_{\rm F} > 0$, where $E_{\rm F} = v_{\rm F}k_{\rm F}$ is the Fermi energy. Here, $v_{\rm F} \sim 10^6$ m/s ($k_{\rm F} = \sqrt{\pi n}$) is the Fermi velocity (Fermi wave number), with $n > 0$ the electron density.

The Dyson equation combined with the approximate $GW$ expression for the electron self-energy define a *self-consistent* approximation, whose self-energy and Green's function can be calculated based on an iterative procedure. One first calculates the self-energy from the $GW$ expression by using in the right-hand side of the *non-interacting* Green's function $G_{\lambda'}(\boldsymbol{k} - \boldsymbol{q}, i\omega_n + i\Omega_m) \to G_{\lambda'}^{(0)}(\boldsymbol{k} - \boldsymbol{q}, i\omega_n + i\Omega_m) = 1/(i\omega_n - \xi_{\boldsymbol{k}-\boldsymbol{q},\lambda'})$. The obtained result is then replaced in the right-hand side of the Dyson equation, obtaining a new Green's function. The latter is then used to re-calculate the self-energy via the $GW$ equation, until self-consistency is achieved. Now, the key point is that deep in the Fermi liquid regime, i.e. for $|\boldsymbol{k}| \simeq k_\mathrm{F}$ and $|\omega|/E_\mathrm{F}, k_\mathrm{B}T/E_\mathrm{F} \ll 1$, the self-energy is a small correction to the bare band energy $\xi_{\boldsymbol{k},\lambda}$ and such self-consistency is unnecessary. In this limit indeed, quasiparticles are long lived because of the ineffectiveness of e-e collisions (Pauli blocking) and $\mathrm{Im}[\Sigma_+(\boldsymbol{k},\omega)] \propto (k_\mathrm{B}T/E_\mathrm{F})^2 + (\omega/E_\mathrm{F})^2$, modulo logarithmic corrections. In this regime, it is therefore well justified to replace $G_{\lambda'}(\boldsymbol{k} - \boldsymbol{q}, i\omega_n + i\Omega_m)$ with $G_{\lambda'}^{(0)}(\boldsymbol{k} - \boldsymbol{q}, i\omega_n + i\Omega_m)$ in the right-hand side of the $GW$ equation obtaining the so-called $G^{(0)}W$ approximation[7,8].

Since this is the simplest possible theory, we use the $G^{(0)}W$ approximation also away from the Fermi liquid regime, being aware of the fact, however, that the lack of full self-consistency is expected to lead to inaccuracies. In particular, it is easy to demonstrate that $\ell_\mathrm{ee}|_{G^{(0)}W} < \ell_\mathrm{ee}|_{GW}$. Since in weakly correlated materials such as graphene the $GW$ approximation is expected to be quantitatively good (i.e. $\ell_\mathrm{ee}|_{GW}$ is expected to be close to the experimentally value of $\ell_\mathrm{ee}$), we do expect the non-self-consistent result $\ell_\mathrm{ee}|_{G^{(0)}W}$ to systematically *underestimate* the experimentally measured $\ell_\mathrm{ee}$. Therefore, in the main text, we have compared experimental data with $\ell_\mathrm{ee}|_{G^{(0)}W}$ after multiplying the latter by a *constant* enhancement factor of 1.3, which is independent of all microscopic parameters (Fig. 2 of the main text).

The quantity $\ell_\mathrm{ee}|_{G^{(0)}W}$ can be calculated numerically once one specifies the dynamically screened potential $W(\boldsymbol{q}, i\Omega_m)$. In the random phase approximation[7], $W(\boldsymbol{q},\omega) = V_{\boldsymbol{q}}/[1 - V_{\boldsymbol{q}}\chi_{nn}^{(0)}(\boldsymbol{q},\omega)]$, where $\chi_{nn}^{(0)}(\boldsymbol{q},\omega)$ is the well-known density-density response function of doped graphene[9] and $V_{\boldsymbol{q}}$ is the 2D Fourier transform of the e-e interaction potential, which is sensitive to screening caused by nearby metal gates and gate dielectrics. For our metal/hBN/graphene/hBN/metal heterostructures, electrostatic calculations yield

$$V_{\boldsymbol{q}} = \frac{4\pi e^2}{q\sqrt{\epsilon_x \epsilon_z}} \frac{\sinh\left(qd\sqrt{\frac{\epsilon_x}{\epsilon_z}}\right) \sinh\left(qd'\sqrt{\frac{\epsilon_x}{\epsilon_z}}\right)}{\sinh\left[q(d+d')\sqrt{\frac{\epsilon_x}{\epsilon_z}}\right]} \tag{S2}$$

where $d'$ ($d$) is the thickness of hBN above (below) graphene, and $\epsilon_x$ and $\epsilon_z$ are the static in-plane and out-of-plane permittivities of hBN. Two metal gates, modelled as perfect conductors, are placed above and below graphene at distances $d'$ and $d \ll d'$, respectively, and are separated from graphene by hBN. Numerical calculations of $\ell_\mathrm{ee}|_{G^{(0)}W}$ have been carried out by using this effective screened e-e interaction for sufficiently large $d' \approx 60$ nm and known $\epsilon_x = 6.70$, and $\epsilon_z = 3.56$ (see, for example, ref. 10). Values of $d, n,$ and $T$ were variables in our calculations. Pertinent results are presented in Fig. 2 of the main text.

For a qualitative understanding of the role of screening, it is useful to obtain an approximate expression for $\ell_\mathrm{ee}|_{G^{(0)}W}$ as a function of all system parameters. To this end, we follow ref. 8 and derive a formula for $\ell_\mathrm{ee}|_{G^{(0)}W}$ which is exact in the Fermi-liquid regime, $k_\mathrm{B}T \ll E_\mathrm{F}$. The calculations follow essentially the same steps as in ref. 8, modulo minor differences, which stem from the regularity of $V_{\boldsymbol{q}}$ in the long-wavelength $q \to 0$ limit and will be discussed elsewhere. Indeed, $\lim_{q \to 0} V_q = 4\pi e^2 d_{\text{eff}}/\epsilon_z \equiv V_0$, where $d_{\text{eff}} = dd'/(d+d')$. This formula allows a simple interpretation. Having the two, top and bottom, gates is like having two capacitors in parallel. Indeed, we can write $V_0 = e^2/C_{\text{eff}}$, where the $C_{\text{eff}} = C_d + C_{d'}$ is the sum of the two relevant geometrical capacitances (per unit area), $C_d = \epsilon_z/(4\pi d)$ and $C_d = \epsilon_z/(4\pi d')$. After restoring $\hbar$, we obtain

$$\lim_{\frac{k_B T}{E_F} \to 0} \ell_{\text{ee}}|_{G^{(0)}W} = \frac{4\hbar v_F E_F}{\pi} \frac{1}{(k_B T)^2 \ln\left(\frac{2E_F}{k_B T}\right)} \left(\frac{1 + 2d_{\text{eff}} q_{\text{TF}}}{2d_{\text{eff}} q_{\text{TF}}}\right)^2 \tag{S3}.$$

Eq. 2 in the main text is simply obtained from Eq. S3 by taking the limit $d' \to \infty$.

Before concluding this section, let us comment on possible corrections to our model caused by the fact that real gates are not the assumed perfect conductors. The effect of a finite density-of-states in a metallic gate can be estimated using the Thomas-Fermi approximation. It is possible to show that, in this approximation, the previous asymptotic result for $\ell_{\text{ee}}|_{G^{(0)}W}$ in the limit $k_B T \ll E_F$ holds if one replaces $d \to d + 1/q_{\text{TF}}$, where $q_{\text{TF}}$ is the Thomas-Fermi screening wavenumber in gate's material. As a crude estimate for graphite, we take $q_{\text{TF}}^{(G)}$ to be the same as that of a three-dimensional metal[7], i.e. $q_{\text{TF}}^{(G)} = \sqrt{4e^2 m_G k_F^{(G)}/(\pi \hbar^2 \epsilon_G)}$, where $m_G \approx 0.2\, m_e$ is the effective mass of charge carriers in graphene, $\epsilon_G \approx 3$, and $k_F^{(G)} = (3\pi^2 n_G)^{1/3}$, with $n_G \approx 10^{19}$ cm$^{-3}$. This yields an extra gate spacing of about 9 Å. This is probably an overestimate as undoped graphene layers provide efficient screening at the same distance due to self-doping[11]. For the other metallic substrates, we find $1/q_{\text{TF}} \approx 2$ Å, that corresponds to interatomic distances as expected.

### #5. Suppression of umklapp e-e scattering by proximity screening

It has been shown[12] that umklapp e-e scattering ($U_{\text{ee}}$) substantially increases the resistivity of high-quality graphene-on-hBN superlattices (SL) in the range of $T$ between 50 and 200 K. The SL potential is generated by the moiré pattern that has a period $\lambda \approx 15$ nm for a perfectly aligned graphene and hBN crystals. $U_{\text{ee}}$ is a process where a crystal lattice (superlattice in our case) provides interacting electrons with an additional momentum kick such that the momentum conservation takes the form $\mathbf{k_3} + \mathbf{k_4} = \mathbf{k_1} + \mathbf{k_2} + \mathbf{g}$, where $\mathbf{k_{1,2}}$ and $\mathbf{k_{3,4}}$ are the initial and final momenta of two electrons near the Fermi level, and $\mathbf{g} = (g_x, g_y)$ is a reciprocal vector of the crystal (Supplementary Fig. 4a). Such a process becomes possible only for $4k_F > g$, where $g = |\mathbf{g}| = \frac{4\pi}{\sqrt{3}\lambda}$ is the length of one of the 6 shortest vectors of the reciprocal SL.

The contribution of $U_{\text{ee}}$ towards graphene's resistivity $\rho$ is given by[12]

$$\Delta \rho = \frac{\hbar \pi}{e^2 k_F} l_{U\text{ee}}^{-1} \text{ with } l_{U\text{ee}}^{-1} = \frac{(k_B T)^2}{12\pi^2 v_F^4 k_F} \sum_{\mathbf{g}} (g_x)^2 \int \frac{d\theta_{\mathbf{k_1}} d\theta_{\mathbf{k_3}}}{|\sin(\theta_{\mathbf{k_2}} - \theta_{\mathbf{k_4}})|} \left| \sum_{i=\text{I}}^{\text{IV}} \sum_{s'=\pm} M_{ss'}^{(i)} \right|^2 \tag{S4}$$

where $\theta_{\mathbf{k}}$ denotes an angle between $\mathbf{k}$ and $x$-axis, $s = \pm$ stands for the conductance/valence-band states (fixed by doping), and $s'$ marks virtual intermediate states. In Eq. S4, the inverse umklapp scattering length, $l^{-1}_{U_{ee}}$, is determined by the sum of four Feynman diagrams shown in Supplementary Fig. 4b, each described by the scattering amplitude $M^{(i)}_{ss'}$ ($i = \text{I, II, III, IV}$). For example, the first diagram gives a contribution

$$M^{(\text{I})}_{ss'} = \frac{W(\mathbf{g})\frac{1+ss'e^{i\theta_{\mathbf{k_1+g}}-i\theta_{\mathbf{k_3}}}}{2}V(|\mathbf{k_2}-\mathbf{k_4}|)\frac{1+e^{i\theta_{\mathbf{k_2}}-i\theta_{\mathbf{k_4}}}}{2}}{sv|\mathbf{k_1}|-s'v|\mathbf{k_1+g}|} \quad (S5)$$

where $W(\mathbf{g})$ stands for the scattering amplitude of an electron off the moiré SL[13,14], and

$$V(q) = \frac{V_q(q,d,d')}{1+V_q(q,d,d')\Pi(q)} \quad (S6)$$

is the Coulomb interaction screened by both gate and the Fermi sea in graphene; $\Pi(q \leq 2k_\text{F}) = \frac{2k_\text{F}}{\hbar\pi v_\text{F}}$ is the Thomas-Fermi polarization operator[15-18]. From the form of $V_q$ in Eq. S2, it is straightforward to see that, for e-e scattering with the momentum transfer $q \sim g/2$, the gate starts playing a notable screening role only if $d_\text{eff} \lesssim \sqrt{\frac{\epsilon_z}{\epsilon_x}}\frac{1}{g} \approx 0.1\lambda$ ($d_\text{eff} \approx d \ll d'$). Expressions for the other diagrams in Supplementary Fig. 4b can be obtained by changing input momenta and $\mathbf{q}$ in Eq. S5.

The $U_{ee}$ contribution, computed using the same SL parameters as those in refs. 12 and 19, exhibits a significant suppression for $d \lesssim 2$ nm (Supplementary Fig. 4c). In these calculations, the absolute value of $\Delta\rho \propto l^{-1}_{U_{ee}}$ obviously depends on the moiré potential's strength. To compare the effect of proximity screening on $U_{ee}$, without relying on a detailed choice of SL parameters, we also plot the ratio $l^{-1}_{U_{ee}}(\infty)/l^{-1}_{U_{ee}}(d)$ at $n \approx -\frac{1}{2}n_0$ and compare the theoretical results with the experimentally found ratio $\Delta\rho(\infty)/\Delta\rho(d)$ [see Fig. 3 of the main text].

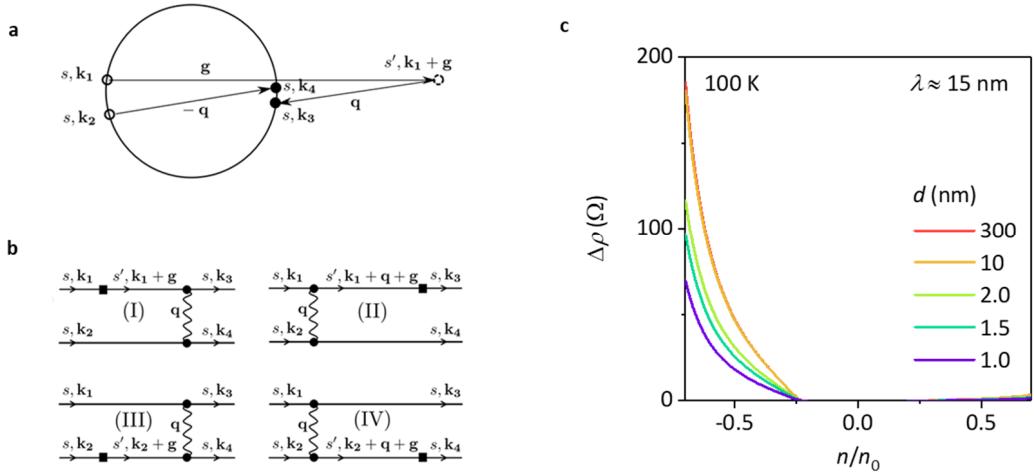

**Supplementary Figure 4 | Screened umklapp e-e scattering in graphene superlattices. a,** Kinematics of $U_{ee}$ scattering. **b,** Feynman diagrams for $M^{(i)}_{ss'}$. **c,** Additional resistivity caused by $U_{ee}$ for different distances to the gate (color coded).